\documentclass[twocolumn,showpacs,preprintnumbers,amsmath,amssymb]{revtex4}
\usepackage{graphicx}
\usepackage{dcolumn}
\usepackage{bm}

\begin{document}

\title{Dependence of transport through carbon nanotubes on local Coulomb
potential.}
\author{A.A.~Zhukov}
\affiliation{Institute of Solid State Physics, Russian Academy of
Science, Chernogolovka, 142432 Russia}

\date{\today}

\begin{abstract}

In this paper, we present the results of helium temperature
transport measurements through carbon nanotubes using an AFM
conductive tip as a mobile gate.  In semiconducting nanotubes we
observe shifting of the conductance peaks with changing of the AFM
tip position. This result can be explained with a particle in the
box quantum model. In high quality metallic nanotubes we observe
that the local Coulomb potential does not destroy the fourfold
degeneracy of the energy levels.

\end{abstract}

\maketitle

The electronic properties of carbon nanotubes have been of
interest since their discovery \cite{ref1}. Quantum dots made from
carbon nanotube have a large energy scale due the quantization
around the nanotube and the additional quantization along the
nanotube length. The fourfold degeneracy of energy levels is
expected from the band structure \cite{ref1a} of metallic
nanotubes. This degeneracy can be observed in the grouping of
conductance peaks in four-peaks clusters \cite{ref2,ref3}.
Nanotubes that have defects or charge impurities are not expected
not to show four-fold degeneracy. Two peaks clusters of spin
degenerate energy states have been found in suspended
semiconductor nanotubes \cite{ref4}.

One of the most powerful methods to observe the influence of
defects on nanotube conductivity is scanned gate microscopy.
\cite{ref5,ref5a}. Until now, most scanned gate microscopy
experiments have been performed at room temperatures \cite{ref5}.
Only one experiment has been reported at low temperatures
($T=0.6$~K) \cite{ref5a}, no degeneracy of energy levels in
conductance peaks was reported.

In this paper we present the results of our low temperature
scanned gate microscopy (SGM). Specifically, we tuned the local
Coulomb potential with our conductive AFM tip while we carried out
transport measurements through semiconducting, and high quality
metallic, carbon nanotubes at $T=4.5$~K. Our results will
concentrate on conductance peaks positions and the four-fold
degeneracy of energy levels in the presence of an external
potential.

The nanotubes used in this experiment were grown by a CVD method
using CO as the feedstock gas. The details of this process are
described elsewhere \cite{ref6}. The contacts to the nanotubes
were defined with e-beam lithography. The semiconducting nanotube,
(Sample~1), was connected with evaporated Cr/Au contacts, and, the
metallic nanotube, (Sample~2), was connected with evaporated Pd/Au
contacts. Additionally, search patterns around the sample were
made to aid in the location of the tube at helium temperatures. An
AFM image taken at 77K of a typical device is shown in Fig.~1a.
Fig.~1b illustrates the general measurement setup of experiment.
The AFM used in our experiment is home made, with a mechanical
positioning range of approximately (1mm)$^2$ and a scanning range
of $(2.5\mu$m$)^2$ at helium temperatures. All results shown in
this paper were measured at $T=$4.5~K.

The insert in Fig.~2a shows a SGM image of Sample~1, obtained at
$V_{BG}$=0~V and $V_t=-2.7$~V, the dark spots represent regions of
higher conductivity. The line parallel to the tube shows the path
of tip movement. The distance to the tube was maintained at a
constant ($d=1.0 \mu$m). Fig.~2a shows dependence of the
conductivity of semiconducting nanotube on back gate voltage
($V_{BG}$), $V_t=-2.7$~V. The differential conductance was
measured by standard AC technique with an excitation voltage of
0.1~mV RMS. Each curve was obtained at a different tip position
(as shown in the insert of Fig.~2a). The results have been
shifted, (by less then $V_{BG}=$0.1~V), to make the conductance
peaks below -3.75~V coincide. The first conductance peaks of holes
in the valence band start to appear at $V_{BG}<-2.25$~V. It can be
seen that for the first few hole states the position of Coulomb
blockade peaks depends on tip position while for the back gate
values below -3.75~V tip displacement is not so significant (ie.
the gaps between the peaks do not change).

Fig.~3 shows a scanned gate image of a high quality metallic tube,
$V_t=-8$~V, $V_{BG}=0$~V. It is easy to see that the equipotential
lines show fourfold grouping when the tip is far from the
nanotube. To ascertain the influence of the local Coulomb
potential close to the nanotube we measured differential
conductance versus back gate voltage while moving tip along an
equipotential line, the dotted line in Fig.~3. The results of this
measurements can be seen in Fig.~4. The differential conductance
was measured by a standard AC technique with an excitation voltage
of 1~mV RMS. The conductance peaks positions do not change and
their grouping in four peak clusters can be seen for all ten
curves, although the heights of the peaks change dramatically.
Thus, we seen no effect on the fourfold energy level degeneracy
due to the local Coulomb potential.

The energy levels shifts in the case of the semiconducting
nanotube can be explained qualitatively in the framework of the
"particle in a 1-D box" model. Half the wavelength of the hole
wave function is equal to $l/n$ where $l$ is the length of the
tube and $n$ is the state number position. When the first holes
fill the nanotube nodes of their wave functions shift considerably
($\delta l \sim d$ where $d$ is distance between tube and line of
tip movement, see Fig.~2a insertion) for each filling state. This
results in measurable deviation of the energy gap in between
conductance peaks for states with low $n$.

We perform calculations of shifting of energy levels due to tip
displacement. We assume that our tip has a shape of ball with
radius of $r_t=100$~nm and potential of $V_t=-2.7$~V. Including
screening from doped back gate placed $\delta$z=1$\mu$m beneath
the nanotube we can find
\begin{eqnarray}
U_{state}(n,x_0)={2 V_t r_t \beta \over
\varepsilon_1+\varepsilon_2} \int_0^l dx sin^2(k_nx)\times \nonumber \\
\times \left[ {{1 \over {\sqrt {(x_0-x)^2+d^2}}}-{1 \over {\sqrt
{(x_2-x)^2+d^2+4\delta z^2}}}} \right],
\end{eqnarray}
where
$\varepsilon_1=1$, $\varepsilon_2=3.9$ are permittivity constants
of Helium and SiO$_2$ substrate correspondingly,
$\beta=c_t/c_{BG}=0.7$ is geometric factor ($c_t$ is capacitance
in between tip and nanotube and $c_{BG}$ is capacitance in between
tip and back gate), $k_n=\pi n/l$ is wave vector of electron or
hall wave function. Additional spacial variation of $U_{state}$
induced by metallic contacts of the nanotube is not taken into
account. Calculated values of energy levels shifts due to charged
tip movement (see Fig.~2b) are comparable with experimental data
(Fig.~2a) if we assume distance in between tip and nanotube
$d=0.3~\mu$m which is significantly smaller than we have in
experiment. The reason of such discrepancy is not clear so only
the general behavior of shifting of calculated energy levels
positions coincide with ones observed experimentally.


In the case of a metallic nanotube all states within the back gate
voltage under investigation are low-lying states ($n$ is large).
This partly resembles the situation with low-lying states in
semiconducting nanotube so survival of 4-peaks clusters at first
glance looks rather reasonable. But the local Coulomb potential
not only changes the potential of the nanotube itself but also the
opacity of the potential barriers in between nanotube and
contacts. This is clearly visible in Fig.~3 as the Coulomb peaks
heights change. So we see that the deviation of coupling strength
of tube to the contacts does not effect the stability of the
fourfold degenerated states.

The influence of Coulomb impurities, defects of the carbon
nanotube and the role of the ends of nanotube have been studied in
the theoretical work of San-Huang Ke {\it et al.} \cite{ref7}. In
this paper the authors performed numerical calculations using
density-functional theory of additional energy spectra for quantum
dots made of ideal nanotubes of various length, nanotubes with
(5-7-7-5) defects and ideal nanotube with Coulomb impurities. This
paper suggests that in an ideal nanotube, Coulomb impurities are
insignificant and do not change additional energy and thus, do not
effect fourfold degeneracy. These theoretical predictions are in
good agreement with our experimental data for low lying states in
high quality metallic nanotube. From the experimental data
presented in Fig.~3 and 4 we can conclude, following paper
\cite{ref7}, that only the imperfections or defects of tube itself
(such as 5-7-7-5) can destroy fourfold degeneracy of the energy
states in metallic carbon nanotubes.

In conclusion, we performed transport measurements of
semiconducting and metallic carbon nanotubes with external tunable
Coulomb potential created by a conductive AFM tip. Shifting of the
conductance peaks positions for the first hole states is observed
and explained qualitatively in framework of particle in 1-D box
model for semiconducting nanotubes. Stability of the fourfold
degeneracy of the energy states in a metallic nanotube to external
Coulomb potential and coupling to the contacts is demonstrated.
These observations are in agreement with previous theoretical
calculations where no influence of local Coulomb potential and the
opacity of potential barriers on degenerate states has been found.
Thus, we can conclude that destruction of the fourfold
degeneration only comes from defects and imperfections of the
metallic nanotube itself.

We thank A. Makarovsky for help with samples preparation.

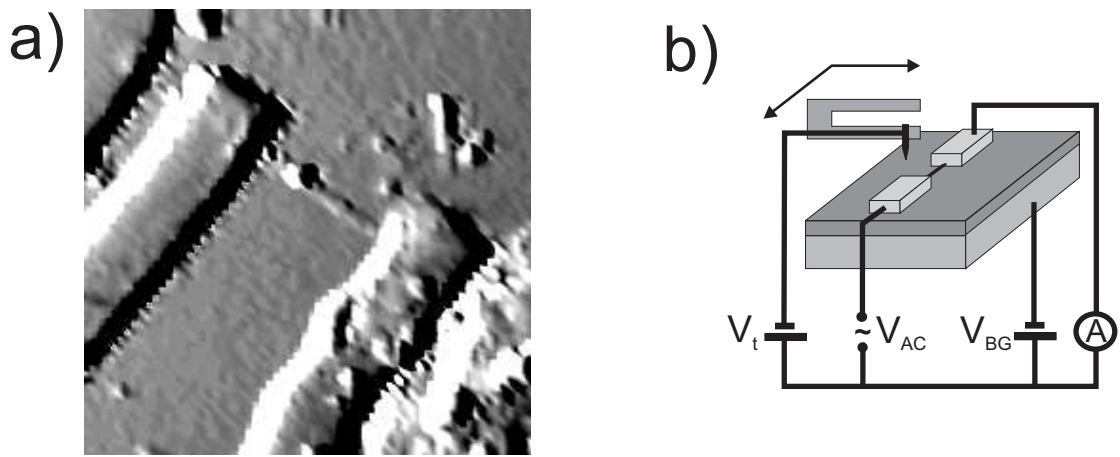
Figure 1: a. AFM image of typical device used in experiment made
at $T=77$~K; b. Electrical scheme of experiment, $V_{BG}$ - back
gate voltage, $V_t$ - tip voltage \cite{ref8}.

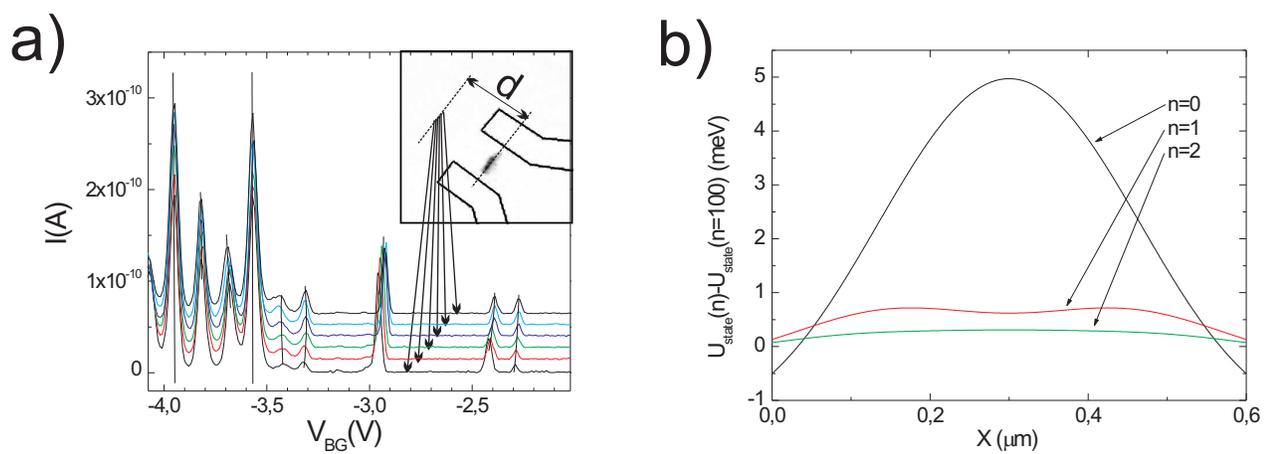
Figure 2: a. Conductivity vs back gate voltage of semiconducting
tube, $V_t=-2.7$~V. Insertion: scanning gate image of Sample~1,
obtained at $V_{BG}=0$~V and $V_t=-2.7$~V, conductive regions are
dark. Distance between tube and line of AFM tip movement $d=1.0
\mu$m, length of tube $l \simeq 0.6 \mu$m. b. Calculated of shifts
of peak positions for $n=$0, 1, 2, shift of energy level for
$n=$100 is subtracted, distance between tube and line of AFM tip
movement $d=0.3 \mu$m. Color online.

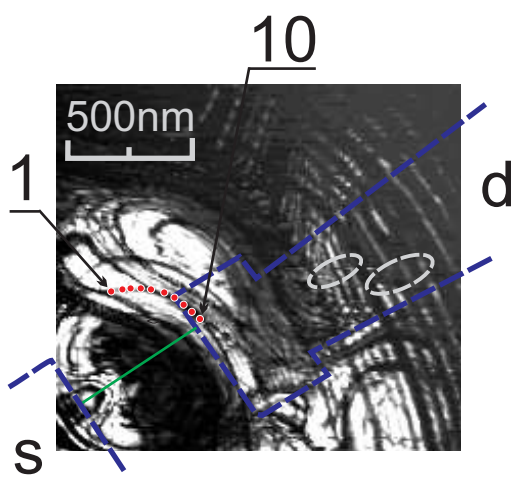
Figure 3: Scanning gate image of high quality metallic nanotube.
Back gate voltage $V_{BG}=0$~V, tip voltage $V_t=-8$~V. More
conductive regions are bright. Circles mark 4-line groups. Dots
are positions 1 to 10 of tip for conductivity vs back gate
measurements, see Fig.~4. Color online.

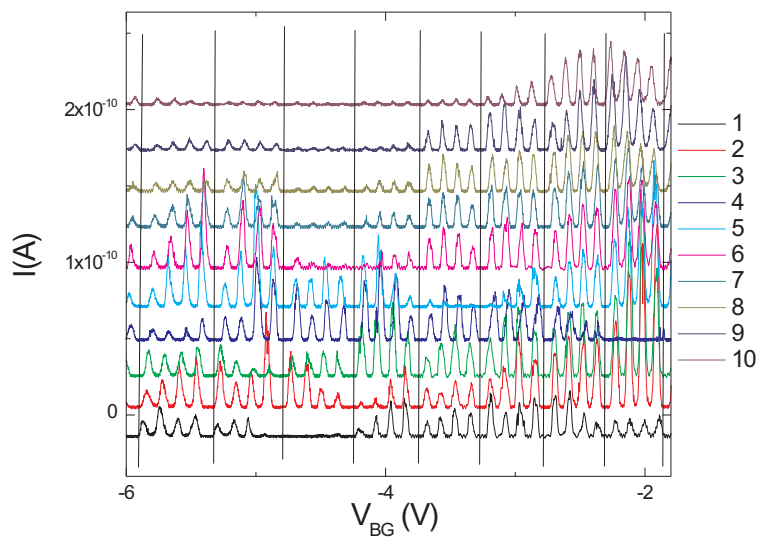
Figure 4: Set of curves 1-10 of conductivity vs back gate voltage,
$V_t=-8$~V. Tip positions are marked in Fig. 3. Color online.



\begin{references}
\bibitem{ref8}
$V_t$ is not actually the voltage applied to the tip. We use
conductive tip made of tungsten wire and it is known that the end
of the tip is covered by oxide which could make an additional
correction of the tip potential. So the value of $V_t$ we
presented is the real potential obtained experimentally from
Kelvin probe measurements (Sample~2) or extracting this value from
additional set of experiments (Sample~1).
\end{references}
\end{document}